\documentclass[12pt,a4paper,twoside]{article}
\date{}

\usepackage{array}
\usepackage{amsmath}
\usepackage{amssymb}
\usepackage{amsthm}
\usepackage[spanish,english]{babel}
\usepackage{booktabs}
\usepackage{float}
\usepackage[a4paper,top=1.5in,bottom=1.5in,inner=1.5in,outer=1.5in]{geometry}
\usepackage{graphics}
\usepackage[dviwin]{graphicx}
\usepackage[latin1]{inputenc}
\usepackage{longtable}
\usepackage{multirow}
\usepackage[round]{natbib}
\usepackage[doublespacing]{setspace}
\usepackage[xllnames,table]{xcolor}
\theoremstyle{definition}

\newtheorem{ejemplo}{Example}

\newtheorem{result}{Result}
\numberwithin{equation}{section}
\numberwithin{table}{section}
\title{A method to find an efficient and robust sampling strategy under model uncertainty}
\author{Edgar Bueno and Dan Hedlin\footnote{Edgar Bueno, Department of Statistics, Stockholm University, 106 91 Stockholm, Sweden, edgar.bueno@stat.su.se; Dan Hedlin, Department of Statistics, Stockholm University, 106 91 Stockholm, Sweden, dan.hedlin@stat.su.se}}

\begin{document}
\maketitle

\begin{abstract}
We consider the problem of deciding on sampling strategy, in particular sampling design. We propose a risk measure, whose minimizing value guides the choice. The method makes use of a superpopulation model and takes into account uncertainty about its parameters. The method is illustrated with a real dataset, yielding satisfactory results. As a baseline, we use the strategy that couples probability proportional-to-size sampling with the difference estimator, as it is known to be optimal when the superpopulation model is fully known. We show that, even under moderate misspecifications of the model, this strategy is not robust and can be outperformed by some alternatives.

\end{abstract}

\section{Introduction}\label{introduction}
We consider the problem of choosing strategy, in particular the design, for the estimation of the total of a study variable in a finite population when a set of $J$ auxiliary variables is available in a list sampling frame. We focus on the estimation of the total.% The type of population we consider is one where study variables form skewed distributions, for example values of properties in a city or turnover of businesses in a country. Such populations call for Poisson sampling, probability proportional-to-size designs or stratified sampling with strata defined by some size measure, which ideally should be measure of the variability with which units contribute to the variance of the estimator(s). In practice, we look for a variable that is correlated with important study variables. We do not focus on Poisson sampling, due to its disadvantage of having a random sample size.  

The decision about sampling strategy involves parameters which are unknown at the stage when the decision needs to be taken. After data collection the parameters can be estimated, although sometimes only under some assumptions. In practice, we often use data from previous waves of a repeated survey, frame variables or data from another survey that is similar to the one that under a planning stage. There is a risk that the available data do not give reliable information about relevant parameters. The method presented here involves a risk measure, which takes into account the possibility of being misled by inaccurate or incorrect beliefs about the values of the needed parameters. The risk measure is derived for the difference and the generalized regression estimators. Other than that, the measure is general. This measure and the discussion of its practical use are the main result of this paper.

One aim when selecting and devising the sampling strategy is efficiency in terms of small mean-squared error. The definition of ``efficiency'' is not unique, however, as it depends on the inference approach. Under the design-based approach, \cite{godambe55}, \cite{lanke73} and \cite{cassel77} show that there is no uniformly best linear estimator, in the sense of being best for all populations. There is no best design either. Therefore, a traditional approach for defining the strategy has been to assume that the finite population is a realization of some superpopulation model. The strategy is then defined in such a way that it minimizes the model expected value of the design mean-squared error, a parameter called anticipated mean-squared error. The adjective ``anticipated'' was first introduced by \citet{isaki82} to emphasize the fact that this is a conceptual mean-squared error which is calculated in advance to sampling, based only on information available prior to sampling.

Assuming that a superpopulation model holds and its parameters are known, several authors have shown that the optimal strategy should make use of a probability proportional-to-size sampling design (e.g. \citealp{hajek59}; \citealp{cassel76}; \citealp{nedyalkova08}). In practice, however, there is not even a consensus about the existence of a generating model, let alone what model to rely on. And even if there is a model, its parameters are unknown. There is evidence, rather empirical, that probability proportional-to-size sampling is not robust towards model misspecifications (e.g. \citealp{holmberg01}). A second result of this paper is to provide some theoretical evidence of this fact.

Many articles discuss robustness in the survey sampling field. \citet{beaumont13}, for instance, propose a robust estimator that downweights influential observations; \citet{royall73} consider robustness under polynomial models; \citet{bramati12} and \citet{zhai15} propose robust stratification methods. We provide theoretical evidence of lack of robustness of proportional-to-size sampling and propose a method for assisting in the decision about the sampling design. 

The contents of the paper are arranged as follows. The optimal strategy under the superpopulation model is defined in section \ref{sec:optimality}. The lack of robustness of this strategy when the model is misspecified is studied in section \ref{sec:robustness}. The method for assisting on the choice of the sampling design is presented in section \ref{sec:decision}. In section \ref{sec:greg}, the risk measure introduced in the previous section is extended to be used together with the GREG estimator. Section \ref{sec:examples} presents numerical illustrations of the results in the paper. First, we illustrate the lack of robustness of probability proportional-to-size sampling and the flexibility of the GREG estimator with a small simulation study. Second, we illustrate the implementation of the risk measure with real survey data. Finally, section \ref{sec:conclusions} presents some conclusions.

\section{Optimal strategy under the superpopulation model}\label{sec:optimality}
Let $U$ be a finite population of size $N$ with elements labeled $\{1,2,\cdots,k,\cdots,N\}$. Let $x_{k}=(x_{1k},x_{2k},\cdots,x_{Jk})$ be a known vector of values of $J$ auxiliary variables and $y_{k}$ the unknown value of a study variable associated to unit $k\in U$. We are interested in the estimation of the total of $y$, $t_{y} = \sum_{U}y_{k}$.% (Note that we have abandoned the use of bold text for denoting vectors. We hope this will not cause any confusion.)

Let $\Omega$ be the power set of $U$. A {\it sample} is any subset $s\in\Omega$ and a {\it sampling design} is a probability distribution on $\Omega$, denoted by $P(S=s)$ or simply $p(s)$. Let $\pi_{k}=\sum_{s\ni k}p(s)$ be the {\it inclusion probability} of $k$ and $\pi_{kl}=\sum_{s\supset\{k,l\}}p(s)$ the {\it joint inclusion probability} of $k$ and $l$. A {\it probability sampling design} is a sampling design such that $\pi_{k}>0$ for all $k\in U$.

An {\it estimator} is a real valued function of the sample, $\hat{t}_{y} = \hat{t}_{y}(S)$. By {\it strategy} we refer to the couple sampling design and estimator, $(p(\cdot),\hat{t}_{y})$.

We consider only probability sampling designs with fixed sample size. As a convenient stepping stone we begin by considering unbiased linear estimators of the form
\begin{equation}\label{equ:estdif}
\hat{t}_{y} = \left(\sum_{U}z_{k}-\sum_{s}\frac{z_{k}}{\pi_{k}}\right)+\sum_{s}\frac{y_{k}}{\pi_{k}} = \sum_{U}z_{k} + \sum_{s}\frac{e_{k}}{\pi_{k}}
\end{equation}
with $z_{k}$ arbitrary known constants and $e_{k}=y_{k}-z_{k}$. This estimator is called the {\it difference estimator}. The estimator defined in this way is said to be {\it calibrated} on $z$ as it satisfies $\hat{t}_{z}=\sum_{U}z_{k}$. Note that if $z_{k}=0$ for all $k\in U$ the estimator reduces to $\hat{t}_{y}=\sum_{s}y_{k}/\pi_{k}$, that is, the Horvitz-Thompson estimator \citep{horvitz52}. In later sections we focus on the generalized regression estimator (GREG).

%A design satisfying $\sum_{s}\frac{z_{k}}{\pi_{k}}=\sum_{U}z_{k}$ for all $s\in\Omega$ such that $p(s)>0$ is called {\it balanced}.% Note that a design is balanced on $z$ if and only if the Horvitz-Thompson estimator, $\hat{t}_{z}=\sum_{s}\frac{z_{k}}{\pi_{k}}$, is calibrated on $z$.

The design MSE of the difference estimator is
\begin{equation}\label{equ:dmseest}
\text{MSE}_\text{p}(\hat{t}_{y}) = \text{MSE}_\text{p}\left(\sum_{s}\frac{e_{k}}{\pi_{k}}\right) =  \sum_{U}\sum_{U}(\pi_{kl}-\pi_{k}\pi_{l})\frac{e_{k}}{\pi_{k}}\frac{e_{l}}{\pi_{l}}.
\end{equation}

As mentioned in the introduction, due to the non-existence of an optimal strategy under the design-based approach, often a superpopulation model, $\xi_{0}$, is proposed and we search for an optimal strategy with respect to the {\it anticipated mean-squared error},
\begin{equation}\label{equ:antmse}
\text{MSE}_{\xi_{0}\text{p}}(\hat{t}_{y}) = \text{E}_{\xi_{0}}\text{MSE}_\text{p}(\hat{t}_{y}) = \text{E}_{\xi_{0}}\text{E}_{p}\left((\hat{t}_{y}-t_{y})^{2}\right)
\end{equation}

We may assume that the $y$-values are realizations of the following model, denoted $\xi_{0}$,
%\begin{equation}\label{equ:xi0}
%Y_{k} =  f(x_{k}|\delta_{1})+\epsilon_{k}\qquad\text{with}\qquad\text{E}_{\xi_{0}}\left(\epsilon_{k}\right)=0,\quad\text{V}_{\xi_{0}}\left(\epsilon_{k}\right) = \sigma_{0}^{2}g(x_{k}|\delta_{2})^{2}\quad\text{and}\quad\text{E}_{\xi_{0}}\left(\epsilon_{k}\epsilon_{l}\right)=0\,\,\,\forall k\ne l
%\end{equation}
\begin{multline}\label{equ:xi0}
Y_{k} =  f(x_{k}|\delta_{1})+\epsilon_{k}\qquad\text{with}\\
\text{E}_{\xi_{0}}\left(\epsilon_{k}\right)=0,\quad\text{V}_{\xi_{0}}\left(\epsilon_{k}\right) = \sigma_{0}^{2}g(x_{k}|\delta_{2})^{2}\quad\text{and}\quad\text{E}_{\xi_{0}}\left(\epsilon_{k}\epsilon_{l}\right)=0\,\,\,\forall k\ne l
\end{multline}
where $\delta=(\delta_{1},\delta_{2})$ is a vector of parameters, $f:\Re^{J}\longrightarrow\Re$ and $g:\Re^{J}\longrightarrow\Re^{+}$. Following \citet{rosen00a}, the terms $f(x_{k}|\delta_{1})$ and $g(x_{k}|\delta_{2})>0$ will be called {\it trend} and {\it spread}, respectively. The term trend should not in general be understood in a temporal sense, rather it refers to the development of $y$-values with $x$. 

Note that under $\xi_{0}$, $e_{k}$ in the difference estimator (\ref{equ:estdif}) is a random variable that represents the distance between the value of the study variable and $z_{k}$, i.e.\ $e_{k} = f(x_{k}|\delta_{1})+\epsilon_{k}-z_{k}$. Therefore $\text{E}_{\xi_{0}}e_{k} = f(x_{k}|\delta_{1})-z_{k}$ and $\text{E}_{\xi_{0}}e_{k}^{2} = (f(x_{k}|\delta_{1})-z_{k})^{2}+\sigma_{0}^{2}\,g(x_{k}|\delta_{2})^{2}$. With some algebra, it can be seen from (\ref{equ:dmseest}) and (\ref{equ:antmse}) that the anticipated MSE of the difference estimator becomes
\begin{equation}\label{equ:amseest}
\text{MSE}_{\xi_{0}\text{p}}(\hat{t}_{y}) = \text{MSE}_\text{p}\left(\sum_{s}\frac{f(x_{k}|\delta_{1})-z_{k}}{\pi_{k}}\right) + \sigma_{0}^{2}\left(\sum_{U}\left(\frac{1}{\pi_{k}}-1\right)g(x_{k}|\delta_{2})^{2}\right)
\end{equation}
\citet{nedyalkova08} derive the anticipated MSE in a more general case.
%Note that $z_{k}$ is not necessarily the same auxiliary variable as the explanatory variable in $f(x_{k}$. The statistician may have made a poor choice when picking $z_{k}$ in the estimator.

\citet{tille17} give the anticipated MSE of the Horvitz-Thompson estimator. The second term in (\ref{equ:amseest}) is the Godambe-Joshi lower bound (e.g.\ \citealt{sarndal92}, p.\ 453).

%\subsection*{The optimal strategy}\label{sub:optimal}
The anticipated MSE in (\ref{equ:amseest}) is the sum of two positive terms. It is easy to see that if
\begin{enumerate}
	\item the estimator is calibrated on $z_{k}=f(x_{k}|\delta_{1})$
\end{enumerate}
the first term vanishes and the anticipated MSE equals the Godambe-Joshi lower bound
\begin{equation}\label{equ:amseopt}
\text{MSE}_{\xi_{0}\text{p}}(\hat{t}_{y}) = \sigma_{0}^{2}\left(\sum_{U}\left(\frac{1}{\pi_{k}}-1\right)g(x_{k}|\delta_{2})^{2}\right)
\end{equation}
Furthermore, after imposing the fixed sample size restriction $\sum_{U}\pi_{k}=n$, if
\begin{enumerate}\setcounter{enumi}{1}
	\item the design is such that $\pi_{k}\propto g(x_{k}|\delta_{2})$, denoted $\pi$ps($g$),
\end{enumerate}
the second term is minimized and we obtain
\begin{equation*}\label{equ:amseopt2}
\text{MSE}_{\xi_{0}\text{p}}^\text{opt}(\hat{t}_{y}) = \sigma_{0}^{2}\left(\frac{1}{n}\left(\sum_{U}g(x_{k}|\delta_{2})\right)^{2} - \sum_{U}g(x_{k}|\delta_{2})^{2}\right).
\end{equation*}
Conditions 1 and 2 suggest the specific roles of the design and the estimator in the sampling strategy. The estimator should ``explain'' the trend in the calibration sense of condition 1. The design should ``explain'' the spread. A strategy that satisfies conditions 1 and 2 simultaneously will be called optimal. In the same sense, any estimator and any design satisfying, respectively, condition 1 and 2, will be called optimal.

\section{Robustness under a misspecified model}\label{sec:robustness}
If the finite population is a realization of the superpopulation model (\ref{equ:xi0}), and if $f$, $g$ and $\delta$ were known, then an optimal strategy could be defined. In this section we study the robustness of this strategy when the model is misspecified.

We begin by defining how \textquotedblleft misspecification\textquotedblright shall be understood in this paper. The {\it working model} $\xi_{0}$ reflects the beliefs the statistician has about the relation between the auxiliary variables $x$ and the study variable $y$ at the design stage. We shall assume that a true, unknown model $\xi$ exists. Any deviation of $\xi_{0}$ with respect to $\xi$ is a misspecification of the model. In order to keep the analysis tractable, we limit ourselves to the situation when the working model is of the form (\ref{equ:xi0}) and the true model, $\xi$, is
%\begin{equation}\label{equ:xi}
%Y_{k} = f(x_{k}|\beta_{1})+\epsilon_{k}\qquad\text{with}\qquad\text{E}_{\xi}\left(\epsilon_{k}\right)=0,\quad\text{V}_{\xi}\left(\epsilon_{k}\right)=\sigma^{2}g(x_{k}|\beta_{2})^{2}\quad\text{and}\quad\text{E}_{\xi}\left(\epsilon_{k}\epsilon_{l}\right)=0\,\,\,\forall k\ne l
%\end{equation}
\begin{multline}\label{equ:xi}
Y_{k} = f(x_{k}|\beta_{1})+\epsilon_{k}\qquad\text{with}\\
\text{E}_{\xi}\left(\epsilon_{k}\right)=0,\quad\text{V}_{\xi}\left(\epsilon_{k}\right)=\sigma^{2}g(x_{k}|\beta_{2})^{2}\quad\text{and}\quad\text{E}_{\xi}\left(\epsilon_{k}\epsilon_{l}\right)=0\,\,\,\forall k\ne l
\end{multline}
where $\beta=(\beta_{1},\beta_{2})$ is a vector of parameters, $f$ and $g$ as in (\ref{equ:xi0}) and $\beta\ne\delta$.% The notation $f_{k}^{\beta}=f(x_{k}|\beta_{1})$ and $g_{k}^{\beta}=g(x_{k}|\beta_{2})$ will be used.

\begin{result}
If $\xi_{0}$ is assumed when $\xi$ is the true superpopulation model, the model expected value of the design MSE in (\ref{equ:dmseest}), under the difference estimator satisfying condition 1 above, becomes
%\begin{equation}\label{equ:antmis}
%\text{MSE}_{\xi\text{p}}(\hat{t}_{y}) = \text{MSE}_\text{p}\left(\sum_{s}\frac{f(x_{k}|\beta_{1})-f(x_{k}|\delta_{1})}{\pi_{k}}\right) + \sigma^{2}\left(\sum_{U}\frac{g(x_{k}|\beta_{2})^{2}}{\pi_{k}}-\sum_{U}g(x_{k}|\beta_{2})^{2}\right)
%\end{equation}
\begin{multline}\label{equ:antmis}
\text{MSE}_{\xi\text{p}}(\hat{t}_{y}) = \text{MSE}_\text{p}\left(\sum_{s}\frac{f(x_{k}|\beta_{1})-f(x_{k}|\delta_{1})}{\pi_{k}}\right) \\+ \sigma^{2}\left(\sum_{U}\frac{g(x_{k}|\beta_{2})^{2}}{\pi_{k}}-\sum_{U}g(x_{k}|\beta_{2})^{2}\right)
\end{multline}
\end{result}
The result is proven by noting that $f(x_{k}|\delta_{1})$ takes the role of $z_{k}$ in (\ref{equ:amseest}) and by taking into account that $e_{k} = f(x_{k}|\beta_{1}) - f(x_{k}|\delta_{1}) + \epsilon_{k}$, therefore $\text{E}_{\xi}e_{k} = f(x_{k}|\beta_{1}) - f(x_{k}|\delta_{1})$ and $\text{E}_{\xi}e_{k}^{2} = (f(x_{k}|\beta_{1})-f(x_{k}|\delta_{1}))^{2}+\sigma^{2}g(x_{k}|\beta_{2})^{2}$. As the model is misspecified, we have deliberately avoided the use of the adjective ``anticipated'' in result 1.

Using result 1, it can be seen that for a design that satisfies condition 2 we obtain
\begin{multline}\label{equ:antmis2}
\text{MSE}_{\xi,\pi\text{ps}}(\hat{t}_{y}) = \left(\frac{\sum_{U}g(x_{k}|\delta_{2})}{n}\right)^{2}\text{MSE}_{\pi\text{ps}}\left(\sum_{s}\frac{f(x_{k}|\beta_{1})-f(x_{k}|\delta_{1})}{g(x_{k}|\delta_{2})}\right) +\\ \sigma^{2}\left(\frac{\sum_{U}g(x_{k}|\delta_{2})}{n}\sum_{U}\frac{g(x_{k}|\beta_{2})^{2}}{g(x_{k}|\delta_{2})}-\sum_{U}g(x_{k}|\beta_{2})^{2}\right).
\end{multline}
%and for model-based stratification,
%\begin{multline*}%\label{equ:antmis3}
%\text{MSE}_{\xi,\text{mb}}(\hat{t}) = \text{MSE}_\text{mb}\left(\sum_{h}\frac{N_{h}}{n_{h}}\sum_{s_{h}}(f(x_{k}|\beta_{1})-f(x_{k}|\delta_{1}))\right) +\\ \sigma^{2}\left(\sum_{h}\frac{N_{h}}{n_{h}}\sum_{U_{h}}g(x_{k}|\beta_{2})^{2}-\sum_{U}g(x_{k}|\beta_{2})^{2}\right)
%\end{multline*}
%And the expected design mean-squared error of STSI--HT is
%\begin{equation}\label{equ:antmis4}
%\text{MSE}_{\xi,\text{st}}(\hat{t}) = \text{MSE}_\text{st}\left(\sum_{h}\frac{N_{h}}{n_{h}}\sum_{s_{h}}f_{k}^{\beta}\right) + \sigma^{2}\left(\sum_{h}\frac{N_{h}}{n_{h}}\sum_{U_{h}}(g_{k}^{\beta})^{2}-\sum_{U}(g_{k}^{\beta})^{2}\right)
%\end{equation}
It is now possible to see that, even under a mild misspecification as the one considered here, the so-called optimal strategy is not necessarily optimal anymore, as its MSE (\ref{equ:antmis2}) can be greater than the MSE obtained under other designs (\ref{equ:antmis}).% Consider, for example, the case of correct trend, $f(x_{k}|\beta_{1}) = f(x_{k}|\delta_{1})$, and misspecified spread. In that case we get
%\[ \text{MSE}_{\xi,\text{mb}}(\hat{t}) < \text{MSE}_{\xi,\pi\text{ps}}(\hat{t})\quad\text{if and only if}\quad\sum_{h}\frac{N_{h}}{n_{h}}\sum_{U_{h}}g(x_{k}|\beta_{2})^{2} < \frac{\sum_{U}g(x_{k}|\delta_{2})}{n}\sum_{U}\frac{g(x_{k}|\beta_{2})^{2}}{g(x_{k}|\delta_{2})}, \]
%which happens, for example, when $\text{Cov}(g(x_{k}|\beta_{2})^{2}/g(x_{k}|\delta_{2})\,,\,g(x_{k}|\delta_{2}))<0$ and the strata are defined as suggested by \citet{wright83}, i.e.\ $n_{h} \propto \sum_{U_{h}}g(x_{k}|\delta_{2})$ with $\sum_{U_{h}}g(x_{k}|\delta_{2}) = \sum_{U}g(x_{k}|\delta_{2})/H$.

\section{Guiding the choice of sampling design with the help of a risk measure}\label{sec:decision}
We have seen in section \ref{sec:robustness} that even a simple misspecification of the working model might result in the so-called optimal strategy not being optimal. It is therefore risky to accept a given model as correct without any type of assessment. While most of the information needed for an ``objective'' evaluation of the model is not available at the design stage, it is possible to reach some degree of confidence about the parameters in the working model that allows for comparing a set of designs and make the decision about which one to implement. In this section we propose a method to assist in the choice of the sampling design.

The expected MSE (\ref{equ:antmis}) in result 1 can be viewed as a function of $\beta$ and $\sigma^{2}$, as everything else is available at the design stage. To begin with, let us assume that $\sigma^{2}$ is also known. Then we can write
\begin{multline*}\label{equ:loss}
L_\text{p}(\beta) = \text{MSE}_{\xi\text{p}}(\beta|x,\delta,\sigma) =\\ \text{MSE}_\text{p}\left(\sum_{s}\frac{f(x_{k}|\beta_{1})-f(x_{k}|\delta_{1})}{\pi_{k}}\right) + \sigma^{2}\left(\sum_{U}\frac{g(x_{k}|\beta_{2})^{2}}{\pi_{k}}-\sum_{U}g(x_{k}|\beta_{2})^{2}\right)
\end{multline*}
For any design, $p(\cdot)$, this function can be evaluated at any $\beta$ and it indicates the loss incurred by assuming that $\delta$ is the right parameter when it is, in fact, $\beta$. We can assume a prior distribution on $\beta$, $h(\beta)$, and calculate the risk under $h$,
\begin{equation}\label{equ:risk}
R(\text{p}) = \text{E}_{h}\left(\text{MSE}_{\xi\text{p}}(\beta|x,\delta,\sigma)\right) = \int_{\Theta}h(\beta)\cdot\text{MSE}_{\xi\text{p}}(\beta|x,\delta,\sigma)d\beta,
\end{equation}
where $\Theta$ is the sample space of $\beta$. The design that yields the smallest risk shall be chosen.

In practice, $\sigma^{2}$ is unknown. We propose three ways for dealing with it. The first one is to redefine $\beta$ as $(\beta,\sigma^{2})$ and calculate the risk as above. The second one is to provide some ``guess'' about it. The third one is to take into account that (proof in the appendix)
\begin{equation}\label{equ:sigma}
\sigma^{2} \approx \frac{S_{f,f}}{\overline{g^{2}}}\left(\frac{1}{R_{f,y}^{2}}-1\right)
\end{equation}
where $S_{f,f}=\sum_{U}(f(x_{k}|\beta_{1})-\bar{f})^{2}/N$, $\bar{f}=\sum_{U}f(x_{k}|\beta_{1})/N$, $\overline{g^{2}}=\sum_{U}g(x_{k}|\beta_{2})^{2}/N$ and $R_{f,y}$ is the correlation between $f(x|\beta_{1})$ and $y$. (In example 3 below, we give a more convenient expression in a special case.) Although $R_{f,y}$ is unknown, for repeated surveys we do have some previous knowledge about it. In other cases it is often possible to have some reasonable ``guess'' about it.

It remains to comment on the choice of the prior distribution $h(\beta)$. The choice of the distribution and its parameters is subjective and defined by the statistician. Nevertheless, it should reflect the available knowledge about the model parameter $\beta$. In particular, $h(\beta)$ should be centered around $\beta=\delta$. Its variance should reflect how confident we are about the working model. Note that a full confidence on the working model would be a density with all its mass at $\beta=\delta$, in which case the risk (\ref{equ:risk}) would be minimized by the $\pi$ps design given by condition 2 in section \ref{sec:optimality}.

It might be argued that by introducing $h(\beta)$ an additional source of subjectivity has been added to the choice of the sampling design. The prior may add a certain Bayesian flavor to the process, but note that $h(\beta)$ is only needed for choosing the design. Hence, the inference is still design-based. Furthermore, relying on an assumed model is also subjective in choice of assumption and it does involve a risk. The risk measure in (\ref{equ:risk}) allows for quantification of that risk.

\section{The risk measure under the Generalized Regression Estimator}\label{sec:greg}
The difference estimator (\ref{equ:estdif}) requires that $\delta_{1}$ is fully specified in order to calculate $f(x_{k}|\delta_{1})$, which is undesirable from a practical standpoint. The generalized regression --GREG-- estimator is an alternative that allows for the estimation of all or some of the components of $\delta_{1}$ at the cost of introducing a small bias. In this section we adapt the material in sections 2 to 4 to strategies using the GREG estimator.

We define the generalized regression estimator in a more general way than in \citet{sarndal92} as follows. Let $a_{k}$ ($k=1,\cdots,N$) a weight defined by the statistician and $\delta_{1}=(\delta_{1}^{*},\delta_{1}^{**})$ where $\delta_{1}^{*}$ are fixed and $\delta_{1}^{**}$ are to be estimated. Let also
\[ \hat{\delta}_{1s}^{**} = \text{argmin}_{\delta_{1}^{**}}\sum_{s}\frac{(y_{k}-f(x_{k}|\delta_{1}))^{2}}{a_{k}\pi_{k}} \]
and $\hat{\delta}_{1s}=(\delta_{1}^{*},\hat{\delta}_{1s}^{**})$. The GREG estimator is
\begin{equation}\label{equ:greg}
\hat{t}_{greg} = \left(\sum_{U}f(x_{k}|\hat{\delta}_{1s})-\sum_{s}\frac{f(x_{k}|\hat{\delta}_{1s})}{\pi_{k}}\right)+\sum_{s}\frac{y_{k}}{\pi_{k}}
\end{equation}
An approximation to the design MSE of the GREG estimator is of the form (\ref{equ:dmseest}) with $e_{k}=y_{k}-f(x_{k}|\hat{\delta}_{1U})$ where $\hat{\delta}_{1U}=(\delta_{1}^{*},\hat{\delta}_{1U}^{**})$ and
\[ \hat{\delta}_{1U}^{**} = \text{argmin}_{\delta_{1}^{**}}\sum_{U}\frac{(y_{k}-f(x_{k}|\delta_{1}))^{2}}{a_{k}} \]

\begin{ejemplo} Let us consider the case where $f(x_{k}|\delta_{1}) = \delta_{1,1}x_{1k}^{\delta_{1,J+1}}+\delta_{1,2}x_{2k}^{\delta_{1,J+2}}+\cdots+\delta_{1,J}x_{Jk}^{\delta_{1,2J}}$. Let $\delta_{1}^{*}=(\delta_{1,J+1},\cdots,\delta_{1,2J})$, $\delta_{1}^{**}=(\delta_{1,1},\cdots,\delta_{1,J})'$ and $x_{k}^{\delta}=(x_{1k}^{\delta_{1,J+1}},\cdots,x_{Jk}^{\delta_{1,2J}})$. In this case we obtain
	\[ \hat{\delta}_{1s}^{**} = \left(\sum_{s}\frac{x_{k}^{\delta'}x_{k}^{\delta}}{a_{k}\pi_{k}}\right)^{-1}\sum_{s}\frac{x_{k}^{\delta'}y_{k}}{a_{k}\pi_{k}}\qquad\text{and}\qquad\hat{\delta}_{1U}^{**} = \left(\sum_{U}\frac{x_{k}^{\delta'}x_{k}^{\delta}}{a_{k}}\right)^{-1}\sum_{U}\frac{x_{k}^{\delta'}y_{k}}{a_{k}}. \]
	%In matrix notation, let $X$ and $X^{\delta}$, respectively, the $N\times J$ matrices with elements $x_{jk}$ and $x_{jk}^{\delta_{1,J+j}}$ in the cell $(k,j)$. Let also $A$ the matrix with diagonal elements $a_{k}$ and $Y$ the column vector with elements $y_{k}$. We have $f(x_{k}|\delta_{1})=X^{\delta}\delta_{1}^{**}$ and
	%\[ \hat{\delta}_{1U}^{**} = \left(X^{\delta'}A^{-1}X^{\delta}\right)^{-1}X^{\delta'}A^{-1}Y \]
Letting the exponents $\delta_{1}^{*}=(\delta_{1,J+1},\cdots,\delta_{1,2J}) = (1,\cdots,1)$, we obtain the classical expression of the GREG estimator found in \citet{sarndal92}.
\end{ejemplo}

\begin{ejemplo}
The case with only one auxiliary variable, i.e.\ $f(x_{k}|\delta_{1}) = \delta_{10}+\delta_{11}x_{k}^{\delta_{12}}$ with $a_{k}=1$, $\delta_{1}^{*}=\delta_{12}$ and $\delta_{1}^{**}=(\delta_{10},\delta_{11})'$ is known as the regression estimator. In this case we obtain the well known result that the design MSE can be approximated by expression (\ref{equ:dmseest}) with $e_{k}=y_{k}-f(x_{k}|\hat{\delta}_{1U})$ where $f(x_{k}|\hat{\delta}_{1U})= \hat{\delta}_{10} + \hat{\delta}_{11}x_{k}^{\delta_{12}}$ and
\[ \hat{\delta}_{11} = \frac{N\sum_{U}x_{k}^{\delta_{12}}y_{k}-\sum_{U}x_{k}^{\delta_{12}}\sum_{U}y_{k}}{N\sum_{U}x_{k}^{2\delta_{12}}-\left(\sum_{U}x_{k}^{\delta_{12}}\right)^{2}}\quad \text{and}\quad \hat{\delta}_{10} =\frac{1}{N}\sum_{U}y_{k} - \hat{\delta}_{11}\frac{1}{N}\sum_{U}x_{k}^{\delta_{12}}. \]
\end{ejemplo}

\subsection*{The misspecified model}
Let us consider again the situation where the statistician uses the working model (\ref{equ:xi0}) but the true model is of the form (\ref{equ:xi}) with $\beta_{1}=(\beta_{1}^{*},\beta_{1}^{**})$, where $\beta_{1}^{*}$ is the counterpart of the fixed component $\delta_{1}^{*}$. The following result states a condition under which result 1 is valid for the GREG estimator.

\begin{result}
If $\hat{\delta}_{1U}^{**}$ converges in distribution to some $\delta_{1}^{**}$ then
%\begin{equation}\label{equ:antmsegreg2}
%\text{MSE}_{\xi\text{p}}(\hat{t}_{greg}) \to \text{MSE}_\text{p}\left(\sum_{s}\frac{f(x_{k}|\beta_{1})-f(x_{k}|\delta_{1})}{\pi_{k}}\right) + \sigma^{2}\left(\sum_{U}\frac{g(x_{k}|\beta_{2})^{2}}{\pi_{k}}-\sum_{U}g(x_{k}|\beta_{2})^{2}\right)
%\end{equation}
\begin{multline}\label{equ:antmsegreg2}
\text{MSE}_{\xi\text{p}}(\hat{t}_{greg}) \to \text{MSE}_\text{p}\left(\sum_{s}\frac{f(x_{k}|\beta_{1})-f(x_{k}|\delta_{1})}{\pi_{k}}\right) \\+ \sigma^{2}\left(\sum_{U}\frac{g(x_{k}|\beta_{2})^{2}}{\pi_{k}}-\sum_{U}g(x_{k}|\beta_{2})^{2}\right)
\end{multline}
where $\delta_{1}=(\delta_{1}^{*},\delta_{1}^{**})$.
\end{result}

The result is proven by using the fact that if $X\overset{d}{\to}C_{1}$ and $Y\overset{d}{\to}C_{2}$ then $X\cdot Y\overset{d}{\to}C_{1}\cdot C_{2}$.

\begin{ejemplo}[Continuation of Example 1] Let the working model be as in example 1 and the true model be $f(x_{k}|\beta_{1}) = \beta_{1,1}x_{1k}^{\beta_{1,J+1}} + \beta_{1,2}x_{2k}^{\beta_{1,J+2}} + \cdots + \beta_{1,J}x_{Jk}^{\beta_{1,2J}}$. Let also $\beta_{1}^{*}=(\beta_{1,J+1},\cdots,\beta_{1,2J})$, $\beta_{1}^{**}=(\beta_{1,1},\cdots,\beta_{1,J})'$ and $x_{k}^{\beta}=(x_{1k}^{\beta_{1,J+1}},\cdots,x_{Jk}^{\beta_{1,2J}})$. In this case, $\hat{\delta}_{1U}^{**} \overset{d}{\to} A\beta_{1}^{**}$, where
\[ A=\left(\sum_{U}\frac{x_{k}^{\delta'}x_{k}^{\delta}}{a_{k}}\right)^{-1}\sum_{U}\frac{x_{k}^{\delta'}x_{k}^{\beta}}{a_{k}}, \]
and (\ref{equ:antmsegreg2}) becomes
%\begin{equation}\label{equ:mseexa2}
%\text{MSE}_{\xi\text{p}}(\hat{t}_{greg}) \to \text{MSE}_\text{p}\left(\sum_{s}\frac{(x_{k}^{\beta}-x_{k}^{\delta}A)\beta_{1}^{**}}{\pi_{k}}\right) + \sigma^{2}\left(\sum_{U}\frac{g(x_{k}|\beta_{2})^{2}}{\pi_{k}}-\sum_{U}g(x_{k}|\beta_{2})^{2}\right).
%\end{equation}
\begin{multline}\label{equ:mseexa2}
\text{MSE}_{\xi\text{p}}(\hat{t}_{greg}) \to \text{MSE}_\text{p}\left(\sum_{s}\frac{(x_{k}^{\beta}-x_{k}^{\delta}A)\beta_{1}^{**}}{\pi_{k}}\right) +\\ \sigma^{2}\left(\sum_{U}\frac{g(x_{k}|\beta_{2})^{2}}{\pi_{k}}-\sum_{U}g(x_{k}|\beta_{2})^{2}\right).
\end{multline}
\end{ejemplo}

\begin{ejemplo}[Continuation of Example 2] Let the working model be as in example 2 and the true model be $f(x_{k}|\beta_{1}) = \beta_{10}+\beta_{11}x_{k}^{\beta_{12}}$ with $\beta_{1}^{*}=\beta_{12}$ and $\beta_{1}^{**}=(\beta_{10},\beta_{11})'$. It can be shown that
%	\begin{equation*}\label{equ:aprox}
%	(x_{k}^{\beta}-x_{k}^{\delta}A)\beta_{1}^{**} = \left(\left(x_{k}^{\beta_{12}}-\overline{x^{\beta_{12}}}\right) - \left(x_{k}^{\delta_{12}}-\overline{x^{\delta_{12}}}\right)\frac{S_{\beta,\delta}}{S_{\delta,\delta}}\right)\beta_{11}
%	\end{equation*}
%	where
%	\begin{align*}
%	\overline{x^{\beta_{12}}} &= \frac{1}{N}\sum_{U}x_{k}^{\beta_{12}} & S_{\beta,\delta} &= \frac{1}{N-1}\sum_{U}(x_{k}^{\beta_{12}}-\overline{x^{\beta_{12}}})(x_{k}^{\delta_{12}}-\overline{x^{\delta_{12}}})\\
%	\overline{x^{\delta_{12}}} &= \frac{1}{N}\sum_{U}x_{k}^{\delta_{12}} & S_{\delta,\delta} &= \frac{1}{N-1}\sum_{U}(x_{k}^{\delta_{12}}-\overline{x^{\delta_{12}}})^{2}
%	\end{align*}
	%Therefore
	(\ref{equ:antmsegreg2}) becomes
	\begin{equation}\label{equ:antmisgreg}
	\text{MSE}_{\xi\text{p}}(\hat{t}_{greg}) \to \beta_{11}^{2}\text{MSE}_\text{p}\left(\sum_{s}\frac{v_{k}}{\pi_{k}}\right) + \sigma^{2}\left(\sum_{U}\frac{g(x_{k}|\beta_{2})^{2}}{\pi_{k}}-\sum_{U}g(x_{k}|\beta_{2})^{2}\right)
	\end{equation}
	with
	\begin{equation}\label{equ:vk}
	v_{k} = \left(x_{k}^{\beta_{12}}-\overline{x^{\beta_{12}}}\right) - \left(x_{k}^{\delta_{12}}-\overline{x^{\delta_{12}}}\right)\frac{S_{\beta,\delta}}{S_{\delta,\delta}},
	\end{equation}
and
	\begin{align*}
\overline{x^{\beta_{12}}} &= \frac{1}{N}\sum_{U}x_{k}^{\beta_{12}} & S_{\beta,\delta} &= \frac{1}{N-1}\sum_{U}(x_{k}^{\beta_{12}}-\overline{x^{\beta_{12}}})(x_{k}^{\delta_{12}}-\overline{x^{\delta_{12}}})\\
\overline{x^{\delta_{12}}} &= \frac{1}{N}\sum_{U}x_{k}^{\delta_{12}} & S_{\delta,\delta} &= \frac{1}{N-1}\sum_{U}(x_{k}^{\delta_{12}}-\overline{x^{\delta_{12}}})^{2}
\end{align*}
Note that (\ref{equ:antmisgreg}) does not depend on $\beta_{10}$.
\end{ejemplo}

If the condition for result 2 holds, i.e.\ if $\hat{\delta}_{1U}^{**}$ converges in distribution to some $\delta_{1}^{**}$, and this value can be calculated or approximated at the design stage for any $\beta$, then the risk (\ref{equ:risk}) can be computed.

Clearly, for the model that has been used in examples 1 and 3, the risk (\ref{equ:risk}) can be obtained by means of expression (\ref{equ:mseexa2}). Furthermore, for the particular case developed in examples 2 and 4, where $f(x_{k}|\beta) = \beta_{10} + \beta_{11}x_{k}^{\beta_{12}}$ and $f(x_{k}|\delta) = \delta_{10} + \delta_{11}x_{k}^{\delta_{12}}$, an alternative approximation of $\sigma^{2}$ is (proof in the appendix)
\begin{equation}\label{equ:sigma2}
\sigma^{2} \approx \beta_{11}^{2}F_{0}\qquad\text{with}\qquad F_{0} = \frac{1}{\overline{x^{2\beta_{2}}}}\frac{S_{1,\beta}^{2}}{S_{1,1}}\left(\frac{1}{R_{x,y}^{2}}-\frac{1}{R_{1,\beta}^{2}}\right)
\end{equation}
where
%\[ \overline{x^{2\beta_{2}}}=\frac{1}{N}\sum_{U}x_{k}^{2\beta_{2}}\qquad S_{1,\beta}=\frac{1}{N}\sum_{U}(x_{k}-\bar{x})(x_{k}^{\beta_{12}}-\overline{x^{\beta_{12}}})\qquad S_{1,1}=\frac{1}{N}\sum_{U}(x_{k}-\bar{x})^{2} \]
\[ \overline{x^{2\beta_{2}}}=\frac{1}{N}\sum_{U}x_{k}^{2\beta_{2}} \quad S_{1,\beta}=\frac{1}{N}\sum_{U}(x_{k}-\bar{x})(x_{k}^{\beta_{12}}-\overline{x^{\beta_{12}}}) \quad S_{1,1}=\frac{1}{N}\sum_{U}(x_{k}-\bar{x})^{2} \]
with $|R_{x,y}|\le|R_{1,\beta}|$ and $R_{1,\beta}$ and $R_{x,y}$ are, respectively, the correlation coefficients between $x$ and $x^{\beta_{12}}$ and between $x$ and $y$. The latter is unknown but often some decent guess about it is available.

The approximation of $\sigma^{2}$ in (\ref{equ:sigma2}) is more convenient than (\ref{equ:sigma}) as now we have that (\ref{equ:antmisgreg}) is approximated by
\begin{equation*}\label{equ:antmisgreg2}
\text{MSE}_{\xi\text{p}}(\hat{t}_{greg}) \approx \beta_{11}^{2}\left[\text{MSE}_\text{p}\left(\sum_{s}\frac{v_{k}}{\pi_{k}}\right) + F_{0}\left(\sum_{U}\frac{g(x_{k}|\beta)^{2}}{\pi_{k}}-\sum_{U}g(x_{k}|\beta)^{2}\right)\right]
\end{equation*}
with $v_{k}$ given by (\ref{equ:vk}). This expression depends neither on the intercept $\beta_{01}$ nor the parameter $\sigma$, and the slope $\beta_{11}$ becomes a proportionality constant that can be ignored.

\section{Numerical examples}\label{sec:examples}
In sections \ref{sec:optimality} and \ref{sec:robustness} we have established that the strategy that couples $\pi$ps sampling with the difference estimator is optimal under a superpopulation model, but it is not robust to misspecifications of this model. In subsection \ref{sub:simulation} we present a small Monte Carlo simulation study carried out to illustrate these results by comparing the optimal strategy and three alternatives.

In sections \ref{sec:decision} and \ref{sec:greg} we introduced a measure that allows for quantifying the risk of implementing a sampling design, so allowing to guide the choice of design. In subsection \ref{sub:decision2} we illustrate the use of the risk measure with real survey data.

\subsection{Simulation study under a misspecified model}\label{sub:simulation}
We compare the efficiency and robustness of four strategies through a simulation study. The four strategies to be compared are $\pi$ps together with the difference estimator (which is optimal when the model is correct), $\pi$ps together with the GREG estimator (optimal design), stratified simple random sampling --STSI-- together with the difference estimator (optimal estimator) and STSI together with the GREG estimator.

Our implementation of $\pi$ps makes use of Pareto $\pi$ps \citep{rosen97}. %The inclusion probabilities induced by the method do not exactly coincide with the desired ones, nevertheless the difference between them is negligible \citep{rosen00b}. 
There is a host of other schemes for drawing $\pi$ps samples. Nevertheless, Pareto $\pi$ps is a convenient method with good properties, see for example \citet{rosen00a}.% An approximation to the design MSE is \citep{rosen00a}
%\begin{equation*}\label{equ:varpipsreg}
%\text{MSE}_{\pi\text{ps}}\left(\sum_{s}\frac{e_{k}}{\pi_{k}}\right) \approx \frac{N}{N-1}\left(\sum_{U}e_{k}^{2}(1-\pi_{k})/\pi_{k} - \frac{\sum_{U}e_{k}(1-\pi_{k})}{\sum_{U}\pi_{k}(1-\pi_{k})}\right).
%\end{equation*}

Our implementation of STSI makes use of model-based stratification \citep{wright83}. We consider $H=5$ strata with boundaries defined using the cum$\sqrt{f}$-rule on $g(x_{k}|\delta_{2})$ as in \citealt{sarndal92} (p. 463) and the sample is allocated using Neyman allocation, $n_{h}\propto N_{h}S_{gh}$. Using the cum$\sqrt{f}$-rule may be suboptimal (see \citealt{sarndal92}, p. 464) but the efficiency of stratification by a continuous size variable is fairly insensitive to the exact choice of boundaries. 

We consider only misspecification of the spread. The trend term is of the form $f(x_{k}|\beta_{1}) = \beta_{10}+\beta_{11}x_{k}^{\beta_{12}}$ with $\beta_{10}=1\,000$, $\beta_{11}=1$ and $\beta_{12}=0.75$, $1$ and $1.25$. The true spread is $g(x_{k}|\beta_{2})=x_{k}^{\beta_{2}}$ with $\beta_{2}=0.5$, $0.75$ and $1$. The working spread is $g(x_{k}|\delta_{2})=x_{k}^{\delta_{2}}$ with $\delta_{2}=0.5$, $0.75$ and $1$.

We will use the difference estimator (\ref{equ:estdif}) calibrated on $f(x_{k}|\beta_{1})$. Regarding the GREG estimator, we will fix $\beta_{12}$, whereas the coefficients $\beta_{10}$ and $\beta_{11}$ will be estimated.

The simulation is set out as follows. The population size is $N=5\,000$. The $x$-values are independent realizations from a gamma distribution with shape $\alpha=4/100$ and scale $\lambda=1200$ plus one unit, whereas $y_{k}$ is a  realization from a gamma distribution with shape and scale
\[ \alpha_{k} = \frac{(\beta_{10}+\beta_{11}x_{k}^{\beta_{12}})^2}{\sigma_{0}^{2}x_{k}^{2\beta_{2}}}\qquad\text{and}\qquad\lambda_{k}=\frac{\sigma_{0}^{2}x_{k}^{2\beta_{2}}}{\beta_{10}+\beta_{11}x_{k}^{\beta_{12}}}, \]
where $\sigma^{2}$ was set in such a way that the correlation between $x$ and $y$ is $\rho=0.95$. The design MSE of a sample of size $n=500$ is then computed for each strategy. The process is iterated $B=5\,000$ times.

\begin{table}[ht]
	\centering
	%\rowcolors{2}{white}{gray!20}
	\caption{Efficiency of three strategies as a percentage of the expected MSE of $\pi$ps--dif under a misspecified model}
	\begin{tabular}{rrr|rrrr}\hline
\multicolumn{7}{c}{Correct model}\\ \hline
$\beta_{12}$&$\beta_{2}$&$\delta_{2}$&$\pi$ps--dif&$\pi$ps--GREG&STSI--dif&STSI--GREG\\ \hline
0.75&0.50&0.50&$2.78\cdot10^{5}$&99.9&57.3&57.3\\
0.75&0.75&0.75&$4.82\cdot10^{4}$&99.6&77.9&77.9\\
0.75&1.00&1.00&$1.90\cdot10^{4}$&99.1&83.3&83.3\\
1.00&0.50&0.50&$7.64\cdot10^{6}$&99.9&57.3&57.3\\
1.00&0.75&0.75&$7.20\cdot10^{5}$&99.7&77.9&77.9\\
1.00&1.00&1.00&$2.14\cdot10^{5}$&99.1&83.2&83.3\\
1.25&0.50&0.50&$1.46\cdot10^{8}$&99.9&57.3&57.3\\
1.25&0.75&0.75&$7.85\cdot10^{6}$&99.7&78.0&78.0\\
1.25&1.00&1.00&$1.81\cdot10^{6}$&99.2&83.2&83.3\\ \hline\hline
\multicolumn{7}{c}{Misspecified model}\\ \hline
$\beta_{12}$&$\beta_{2}$&$\delta_{2}$&$\pi$ps--dif&$\pi$ps--GREG&STSI--dif&STSI--GREG\\ \hline
 0.75&0.50&0.75&$3.98\cdot10^{5}$& 99.9& 98.9& 98.9\\
%0.75&0.50&1.00&$7.79\cdot10^{5}$& 99.9&185.6&185.6\\
%0.75&0.75&0.50&$6.91\cdot10^{4}$&100.0& 93.0& 93.1\\
 0.75&0.75&1.00&$6.45\cdot10^{4}$& 99.5&114.5&114.5\\
 0.75&1.00&0.50&$4.73\cdot10^{4}$&100.1&133.9&134.0\\
%0.75&1.00&0.75&$2.46\cdot10^{4}$& 99.7& 88.9& 89.0\\
%1.00&0.50&0.75&$1.09\cdot10^{7}$& 99.9& 98.9& 98.9\\
 1.00&0.50&1.00&$2.14\cdot10^{7}$& 99.9&185.7&185.7\\
 1.00&0.75&0.50&$1.03\cdot10^{6}$&100.1& 93.1& 93.1\\
%1.00&0.75&1.00&$9.65\cdot10^{5}$& 99.5&114.5&114.5\\
%1.00&1.00&0.50&$5.34\cdot10^{5}$&100.2&134.0&134.1\\
 1.00&1.00&0.75&$2.77\cdot10^{5}$& 99.8& 89.0& 89.1\\
 1.25&0.50&0.75&$2.09\cdot10^{8}$& 99.9& 98.9& 98.9\\
%1.25&0.50&1.00&$4.09\cdot10^{8}$& 99.9&185.7&185.7\\
%1.25&0.75&0.50&$1.13\cdot10^{7}$&100.1& 93.1& 93.1\\
 1.25&0.75&1.00&$1.05\cdot10^{7}$& 99.6&114.6&114.6\\
 1.25&1.00&0.50&$4.50\cdot10^{6}$&100.3&134.0&134.2\\
%1.25&1.00&0.75&$2.34\cdot10^{6}$& 99.9& 89.0& 89.1\\
	\end{tabular}
	\label{tab:table1}
\end{table}

Table \ref{tab:table1} shows the results of the simulation study. The first three columns indicate the model parameters. The fourth column shows the (simulated) expected MSE of the strategy $\pi$ps--dif, whereas the last three columns show the (simulated) efficiency of the strategies $\pi$ps--GREG, STSI--dif and STSI--GREG compared to $\pi$ps--dif (as a percentage), with efficiency defined as 
\[ \text{eff} = \frac{1}{B}\sum_{r=1}^{B}\text{eff}^{(r)} \qquad\text{where}\qquad \text{eff}^{(r)} = \frac{\text{MSE}_{\xi,\pi\text{ps}}^{(r)}(\hat{t}_{y})}{\text{MSE}_{\xi,\text{p}}^{(r)}(\hat{t}_{y})}, \]
in such a way that a value of 100 indicates that the strategy is as efficient as $\pi$ps--dif and values smaller (larger) than one indicate that the strategy is less (more) efficient than $\pi$ps--dif.

The upper part of Table \ref{tab:table1} shows the case when the working model coincides with the true model. As expected, the strategy that couples $\pi$ps with the difference estimator ($\pi$ps--dif) was always more efficient than the remaining strategies. Nevertheless, the loss in efficiency due to estimating some parameters through the GREG estimator is negligible. On the other hand, there is a remarkable loss in efficiency due to the use of STSI instead of $\pi$ps. Finally, it is noted from (\ref{equ:amseopt}) that as the anticipated MSE for all strategies does not depend on the trend $f$ but only on the spread $g$, the efficiency remains constant under the same value of $\delta_{2}$, independently of the value of $\beta_{12}$.

The lower part of Table \ref{tab:table1} shows some comparisons under a misspecified model, in particular, a misspecified spread. It can be noted that even under this mild misspecification of the model, $\pi$ps--dif is not necessarily the best strategy anymore as the strategies using STSI were more efficient in several cases. However, it is not evident when will STSI be more efficient than $\pi$ps or vice versa. The risk measure introduced in section \ref{sec:decision} can be used to guide the choice between designs. The results shown in this section agree with those shown by for example \citet{holmberg01}.

\subsection{Using the risk measure for choosing the design in a real survey}\label{sub:decision2}
In this subsection we illustrate the implementation of the risk measure using data from a real survey. We want to estimate $t_{y}=\sum_{U}y_{k}$ where $U$ is the set of residential properties in Bogotá, Colombia (of size $N=681\,276$) and $y_{k}$ is the value of the $k$th property in 2017 in COP. $x_{k}$, the built-up area of the $k$th property in square meters, is known for every $k\in U$. The auxiliary variable $x$ has mean 184, standard deviation 110 and skewness 2.57. The desired sample size is $n=1\,000$.
	
We assume that a model of the type $\xi_{0}$ with $f(x_{k}|\delta_{1})=\delta_{10}+\delta_{11}x_{k}^{\delta_{12}}$ and $g(x_{k}|\delta_{2})=x_{k}^{\delta_{2}}$ adequately describes the association between $x$ and $y$. We plan to use the GREG estimator for estimating $\delta_{10}$ and $\delta_{11}$, i.e.\ $\delta_{1}^{**} = (\delta_{10},\delta_{11})$. We will use the risk (\ref{equ:risk}) in order to assist the decision between $\pi$ps or STSI. $H=6$ strata are used and we take $h(\beta_{12},\beta_{2})$ as a bivariate normal distribution with no correlation between $\beta_{12}$ and $\beta_{2}$. We consider two cases with different degrees of confidence regarding the working model.
	
\paragraph{Case 1.} In this case no information about $\delta_{12}$, $\delta_{2}$ or $R_{x,y}$ is available. Naive values of $\delta_{12}=1$, $\delta_{2}=1$ and $R_{x,y}=0.75$ are considered. In order to reflect the uncertainty, $h(\beta)$ should have a large variance, therefore we set
\[ \left[\begin{matrix}\beta_{12}\\\beta_{2}\end{matrix}\right] \sim \text{N}\left(\left[\begin{matrix}1.0\\1.0\end{matrix}\right],\left[\begin{matrix}0.3295^{2}&0\\0&0.3295^{2}\end{matrix}\right]\right). \]
Evaluation of (\ref{equ:risk}) yields $R(\pi\text{ps})=6.89\cdot10^{15}\beta_{11}^{2}$ and $R(\text{st})=1.59\cdot10^{15}\beta_{11}^{2}$, suggesting that a stratified design should be used.
	
The design MSE of both strategies is computed and we get, $\text{MSE}_{\pi\text{ps}}(\hat{t}_{y}) = 2.29\cdot 10^{25}$ and $\text{MSE}_{\text{st}}(\hat{t}_{y})=1.36\cdot 10^{25}$. The strategy suggested by (\ref{equ:risk}) was indeed the best choice.
	
\paragraph{Case 2.} Using a sample from 2010, prior values of $\delta_{12}=1.9$, $\delta_{2}=2$ and $R_{x,y}=0.7$ are proposed. As the uncertainty here is smaller than that in case 1, we set a smaller variance,
\[ \left[\begin{matrix}\beta_{12}\\\beta_{2}\end{matrix}\right] \sim \text{N}\left(\left[\begin{matrix}1.9\\2.0\end{matrix}\right],\left[\begin{matrix}0.2471^{2}&0\\0&0.2471^{2}\end{matrix}\right]\right) \]
Evaluation of (\ref{equ:risk}) yields $R(\pi\text{ps})=7.08\cdot10^{22}\beta_{11}^{2}$ and $R(\text{st})=4.06\cdot10^{18}\beta_{11}^{2}$, suggesting that a stratified design should be used.
	
The design MSE of both strategies is computed and we get $\text{MSE}_{\pi\text{ps}}(\hat{t}_{y}) = 1.85\cdot 10^{28}$ and $\text{MSE}_{\text{st}}(\hat{t}_{y})=1.91\cdot 10^{25}$. Note that the use of (\ref{equ:risk}) prevented us from using $\pi$ps, whose MSE is almost one thousand times bigger than the one under stratified sampling!

\section{Conclusions}\label{sec:conclusions}
The strategy that couples $\pi$ps with the difference estimator is optimal when the parameters of the superpopulation model are known. Taking into account that these assumptions are seldom satisfied, it was shown in section \ref{sec:robustness} and illustrated in subsection \ref{sub:simulation} that this optimality breaks down even under small misspecifications of the model.
	
In section \ref{sec:decision} we propose a method for choosing the sampling design, which is extended to its use with the GREG estimator in section \ref{sec:greg}. The method allows for taking the uncertainty about the model parameters into account by introducing a prior distribution on them. Although it could be argued that a source of subjectivity is added by introducing a prior distribution on the parameters, our view is that it is more subjective to choose the design without any type of assessment of the assumptions. Furthermore, inference is still design-based, as the prior is used only for choosing the design.
	
The method was illustrated with a real dataset, yielding satisfactory results. It should be noted that although the illustrations used stratified simple random sampling, the method in this article is valid for any sampling design.

\appendix
\section*{Appendix. Proof of (\ref{equ:sigma})}
%\begin{proof}[Proof of (\ref{equ:aprox})] Note that
%\begin{align*}
%	\text{E}_{\xi}\hat{\delta}_{11} =& \frac{N\sum_{U}x_{k}^{\delta_{12}}(\beta_{10}+\beta_{11}x_{k}^{\beta_{12}})-\sum_{U}x_{k}^{\delta_{12}}\sum_{U}(\beta_{10}+\beta_{11}x_{k}^{\beta_{12}})}{N\sum_{U}x_{k}^{2\delta_{12}}-\left(\sum_{U}x_{k}^{\delta_{12}}\right)^{2}}  = \beta_{11}\frac{S_{\beta,\delta}}{S_{\delta,\delta}} \\
%	\text{E}_{\xi}\hat{\delta}_{10} =& \text{E}_{\xi}\left[\frac{1}{N}\sum_{U}Y_{k} - \hat{\delta}_{11}\frac{1}{N}\sum_{U}x_{k}^{\delta_{12}}\right] = \frac{1}{N}\sum_{U}\text{E}_{\xi}Y_{k}-\overline{x^{\delta_{12}}}\text{E}_{\xi}\hat{\delta}_{11} = \beta_{10}-\beta_{11}\left(\overline{x^{\delta_{12}}}\frac{S_{\beta,\delta}}{S_{\delta,\delta}}-\overline{x^{\beta_{12}}}\right)
%\end{align*}
%Therefore
%\[ \text{E}_{\xi}f(x_{k}|\hat{\delta}_{1U}) = \text{E}_{\xi}\left(\hat{\delta}_{10} + \hat{\delta}_{11}x_{k}^{\delta_{12}}\right) = \beta_{10}-\beta_{11}\left(\overline{x^{\delta_{12}}}\frac{S_{\beta,\delta}}{S_{\delta,\delta}}-\overline{x^{\beta_{12}}}\right) + \beta_{11}\frac{S_{\beta,\delta}}{S_{\delta,\delta}}x_{k}^{\delta_{12}} \]
%\end{proof}

\begin{proof}%[Proof of (\ref{equ:sigma})]
	The following expectations are required in the proof,
	\begin{align}
	\text{E}_{\xi}Y_{k} &= \text{E}_{\xi}\left[f(x_{k}|\beta_{1})+\epsilon_{k}\right] = f(x_{k}|\beta_{1})\label{ykapp} \\ 
	\text{E}_{\xi}Y_{k}^{2} &=  \text{E}_{\xi}\left[\left(f(x_{k}|\beta_{1})+\epsilon_{k}\right)^{2}\right] = f(x_{k}|\beta_{1})^{2}+\sigma^{2}g(x_{k}|\beta_{2})^{2} \label{yk2app}
	\end{align}
	$\text{E}_{\xi}\overline{Y}$, $\text{E}_{\xi}\overline{Y^{2}}$ and $\text{E}_{\xi}\overline{fY}$ are obtained using (\ref{ykapp}) and (\ref{yk2app}),
	\begin{align}
	\text{E}_{\xi}\overline{Y} &= \text{E}_{\xi}\left[ \frac{1}{N}\sum_{U}Y_{k}\right] = \frac{1}{N}\sum_{U}\text{E}_{\xi}Y_{k} =  \frac{1}{N}\sum_{U}f(x_{k}|\beta_{1}) \equiv \overline{f} \label{ybarapp} \\
	\text{E}_{\xi}\overline{Y^{2}} &= \text{E}_{\xi}\left[\frac{1}{N}\sum_{U}Y_{k}^{2}\right] = \frac{1}{N}\sum_{U}(f(x_{k}|\beta_{1})^{2}+\sigma^{2}g(x_{k}|\beta_{2})^{2}) \equiv \overline{f^{2}} + \sigma^{2}\overline{g^{2}} \label{y2barapp} \\
	\text{E}_{\xi}\overline{fY} &= \text{E}_{\xi}\left[\frac{1}{N}\sum_{U}f(x_{k}|\beta)Y_{k}\right] = \frac{1}{N}\sum_{U}f(x_{k}|\beta)\text{E}_{\xi}Y_{k} =  \frac{1}{N}\sum_{U}f(x_{k}|\beta)^{2} = \overline{f^{2}} \label{xybarapp}
	\end{align}
	Now, using (\ref{ybarapp}), (\ref{y2barapp}) and (\ref{xybarapp}) we get
	\begin{align}
	\text{E}_{\xi}\left[\overline{fY}-\overline{f}\,\overline{Y}\right] =& \overline{f^{2}} - \overline{f}^{2} = S_{f,f}\label{sxyapp} \\
	\text{E}_{\xi}\left[\overline{Y^{2}}-\overline{Y}^{2}\right] =& \overline{f^{2}} + \sigma^{2}\overline{g^{2}} - \overline{f}^{2} = S_{f,f} + \sigma^{2}\overline{g^{2}} \label{syyapp}
	\end{align}
	Using (\ref{sxyapp}) and (\ref{syyapp}), we obtain an approximation to the correlation coefficient, $R_{f,y}$,
	\begin{equation}\label{r2xy}
	R_{f,y}^{2} = \frac{(\overline{fy}-\overline{f}\overline{y})^{2}}{(\overline{f^{2}}-\overline{f}^{2})(\overline{y^{2}}-\overline{y}^{2})} \approx \frac{\text{E}^{2}_{\xi}\left[\overline{fY}-\overline{f}\,\overline{Y}\right]}{\text{E}_{\xi}\left[(\overline{f^{2}}-\overline{f}^{2})\left(\overline{Y^{2}}-\overline{Y}^{2}\right)\right]} = \frac{S_{f,f}}{S_{f,f} + \sigma^{2}\overline{g^{2}}}
	\end{equation}
	Solving (\ref{r2xy}) for $\sigma^{2}$ we get (\ref{equ:sigma}), as desired.
	The proof of (\ref{equ:sigma2}) is analogous.
\end{proof}

\end{document}